\newif\ifpreprint
\preprintfalse
\ifpreprint
   \documentclass[preprint,showpacs,preprintnumbers,amsmath,amssymb]{revtex4}
\else
   \documentclass[prl,twocolumn,noshowpacs,preprintnumbers,amsmath,amssymb,floatfix]{revtex4}
\fi
\usepackage{graphicx}% Include figure files
\graphicspath{{Figures/}}
\DeclareGraphicsExtensions{.eps,.ps}

\usepackage{dcolumn}% Align table columns on decimal point
\usepackage{bm}% bold math

\begin{document}
\newlength\figwidth
\setlength\figwidth{0.95\columnwidth}

% \title{The Order of Phase Transitions in the Activated Decay of Metastable States}
\title{The Order of Phase Transitions in Barrier Crossing}

\author{J.\ B\"urki}
\author{C.\ A.\ Stafford}%
% \email{stafford@physics.arizona.edu}
\affiliation{%
Physics Department, University of Arizona, 1118 E.\ 4th Street, Tucson, AZ 85721
}%
\author{D.\ L.\  Stein}
\affiliation{Department of Physics and Courant Institute of Mathematical Sciences, New York University, New York, NY 10003}


\date{Submitted February 12, 2008; Accepted by Phys. Rev. E on April 29, 2008}% It is always \today, today,

% \ifpreprint
\begin{abstract}
A spatially extended classical system with metastable states subject to
weak spatiotemporal noise can exhibit a transition in its activation
behavior when one or more external parameters are varied.  Depending on the
potential, the transition can be first or second-order, but there exists no
systematic theory of the relation between the order of the transition and
the shape of the potential barrier.  In this paper, we address that
question in detail for a general class of systems whose order parameter is
describable by a classical field that can vary both in space and time, and
whose zero-noise dynamics are governed by a smooth polynomial potential.
We show that a quartic potential barrier can only have second-order
transitions, confirming an earlier conjecture~\cite{Stein04}.  We then
derive, through a combination of analytical and numerical arguments, both
necessary conditions and sufficient conditions to have a first-order vs.~a
second-order transition in noise-induced activation behavior, for a large
class of systems with smooth polynomial potentials of arbitrary order.  We
find in particular that the order of the transition is especially sensitive
to the potential behavior near the {\it top\/} of the barrier.
\end{abstract}

% \fi

\maketitle

\section{Introduction}
\label{sec:Intro} % Introduction

When a spatially extended classical system with multiple locally
stable states is perturbed by weak spatiotemporal noise, a transition
in its activation behavior can occur as one or more parameters of the
system are varied~\cite{MS01b,MSspie,Stein04}.  In the simplest
one-dimensional systems this parameter is simply the length of the
interval on which the system is defined, but for more complicated
systems other parameters come into play.  For example, a transition
from Arrhenius to non-Arrhenius behavior in thermally activated
magnetization reversal in thin annular nanomagnets can occur either as
ring size increases or as the externally applied magnetic field
decreases~\cite{MSK05,MSK06}.  
Another example is a crossover from uniform to instanton-like decay 
of metastable metal nanowires~\cite{BSS05} as either the length of the 
nanowire or the stress applied to it is varied.  
A similar crossover, from thermal activation to quantum tunneling, 
occurs in various systems as temperature is lowered~
\cite{Goldanskii59,Affleck81,Wolynes81,CL81,LO83,GW84,RHF85,Chudnovsky92,KT97,GB97,FY99}.

These two cases are formally related through a mapping that identifies
interval length (and/or magnetic field, stress, if appropriate) in the
classical field case to temperature in the quantum case; in
particular, increasing interval length in the former corresponds to
the lowering of temperature in the latter, with thermal activation in
a classical field theory of infinite domain size mapping to
zero-temperature tunneling in quantum field theories.  These
transitions are fundamentally different from the more usual sort, in
which a change in order parameter (i.e., expectation value of the
field) results from varying a control parameter, such as coupling
strengths in the Hamiltonian or noise amplitude.  (For an extensive
discussion of this more conventional kind of transition within a noisy
field-theoretical framework, see, for example,~\cite{BL04}.) The extent
to which the more unconventional type of transition under discussion
here can be compared to a true second-order phase transition, and
where the analogy breaks down, was discussed extensively
in~\cite{Stein05}. (For recent work on discretized versions of these
and similar models, see~\cite{BFG1,BFG2}.)

Chudnovsky~\cite{Chudnovsky92} first noted that the
classical$\leftrightarrow$quantum transition in an extended system with a single
degree of freedom can be either first- or second-order, depending on the
potential.  This observation has important applications.  First-order
transitions in the classical-quantum escape rate have been considered in
anisotropic bistable large-spin models~\cite{CG97} and biaxial spin systems
subject to longitudinal fields~\cite{GC99}, and have been more generally
considered in the decay of metastable states in quantum field
theories~\cite{KuTi97,GB97}.  A general discussion can be found
in~\cite{Chudnovskybook}.

Similarly, a first-order transition can also occur in classical transitions
between two thermally activated regimes, as was recently found by the
authors~\cite{BSS05} in the decay of nanowires due to thermal fluctuations
(cf.~Fig.~2 of Ref.~\cite{BSS05}). %~\ref{fig:firstorderDW}).  
Despite its
potential importance, relatively little systematic work has been done to
identify the general conditions under which one or the other kind of
transition occurs.  It was conjectured~\cite{Stein04} that smooth
polynomial potentials with terms no higher than quartic display only
second-order transitions.  It is also known that higher-order terms can
lead to a first-order transition ~\cite{KuTi97}.  At the present time,
however, there is no systematic theory of how the order of the transition
depends on potential characteristics.  The purpose of this paper is to
address that problem.

\section{Model}
\label{sec:model} % Description of model

We will consider a general class of models of extended systems
describable by a classical field $\phi(z,t)$ defined on the spatial
interval $[-L/2,L/2]$, subject to a potential $V(\phi)$ and perturbed
by spatiotemporal white noise~\cite{note}.  Time evolution is governed
by the stochastic Ginzburg-Landau equation
\begin{equation}\label{eq:GL} % Ginzburg-Landau equation
  \partial_t\phi=\partial_{zz}\phi - \partial_\phi V(\phi) +
  \sqrt{2T}\xi(z,t)\, ,
\end{equation}
where all dimensional quantities have been scaled out.
%$\partial_\alpha\equiv\partial/\partial\alpha$.  
The first term on the RHS arises from a field `stiffness', i.e., 
an energy penalty for spatial variations of the field.  
The noise $\xi(z,t)$ satisfies 
$\langle\xi(z_1,t_1)\xi(z_2,t_2)\rangle=\delta(z_1-z_2)\delta(t_1-t_2)$, 
and its magnitude $T$ is small compared to all other energy scales in the
problem (formally, our analysis will be asymptotically valid in the $T\to
0$ limit).

The zero-noise dynamics of~(\ref{eq:GL}) can be written as the variation of
an action ${\cal H}$ with the field $\phi$:
\begin{equation}\label{eq:variation}  % Zero-noise dynamics
 % \dot\phi
 \partial_t\phi = -\delta{\cal H}/\delta\phi
\end{equation}
with
\begin{equation}\label{eq:action} % Effective energy for fluctuations
  {\cal H}[\phi]\equiv \int_{-L/2}^{L/2}dz\ 
    \left[\frac12(\partial_z\phi)^2 +V(\phi)\right] \, .
\end{equation}
Stationary solutions of~(\ref{eq:variation}) describe stable, metastable, and
transition (i.e., saddle) states of the system.

In the weak-noise ($T\to 0$) limit, the classical activation rate for a
transition out of a (meta)stable well is
\begin{equation}\label{eq:Kramers}  % Kramers rate
  \Gamma\sim \Gamma_0\exp(-\Delta W/T)\, ,
\end{equation}
where the activation barrier $\Delta W$ is the action difference between
the transition state and the initial (meta)stable state.  The rate
prefactor $\Gamma_0$ is determined by fluctuations about the most probable
escape path.  
When the top of the barrier is locally quadratic, the
prefactor $\Gamma_0$ is {\it independent\/} of temperature.  
In such circumstances the escape rate~(\ref{eq:Kramers}) is said to be of the
Arrhenius-van't Hoff (or often simply Arrhenius) form.  
Here we will mostly be concerned with the behavior of the activation barrier 
$\Delta W$.

\section{Transition in activation behavior}
\label{sec:transition}  % Transition in activation behavior

\begin{figure}[t] %%%  Figure 1  %%%%%%%%%%%%%%%%%%%%%%%%%%%%%%%%
  \includegraphics[width=0.95\columnwidth]{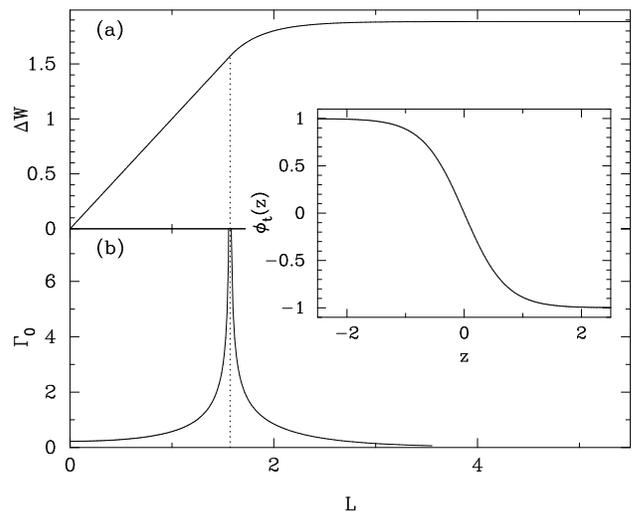} % {activation.eps}
  \caption[The activation energy and prefactor.]{
    (a) The activation energy $\Delta W$ and (b) the rate prefactor $\Gamma_0$ 
    as functions of the interval length~$L$,
    for the potential given by Eq.~(\ref{eq:quartic}) with Neumann boundary 
    conditions.  
    The dashed line indicates the critical interval length $L_c=\pi/2$ at which 
    the saddle state bifurcation takes place, showing the
    power-law divergence of the prefactor.
    The transition  state $\phi_t(z)$ for $L=5$ (corresponding to $m=0.986$) 
    described by  Eq.~(\ref{eq:instanton}) is displayed in the inset 
    (only one of the  symmetric pair is shown.)
    Note that quantities in all figures are expressed in dimensionless units 
    (see text.)
    }
    \label{fig:acten} % Activation energy and prefactor
\end{figure}

We briefly summarize here the derivation of a second-order transition
in the noise-induced barrier crossing described by~(\ref{eq:GL}) between wells
in the simple bistable symmetric quartic potential
\begin{equation}\label{eq:quartic}  % Quartic potential
  V_s(\phi)=(\phi^2-1)^2, 
\end{equation}
with Neumann boundary conditions $\partial_z\phi|_{-L/2} = 
\partial_z\phi|_{L/2} = 0$.  
%$\partial\phi/\partial L|_{-L/2}=\partial\phi/\partial L|_{L/2}=0$.  
The discussion follows that of~\cite{Stein05}, to which we refer 
the reader for details.

\begin{figure}[t] %%%  Figure 2  %%%%%%%%%%%%%%%%%%%%%%%%%%%%%%%%
  \includegraphics[width=0.99\columnwidth]{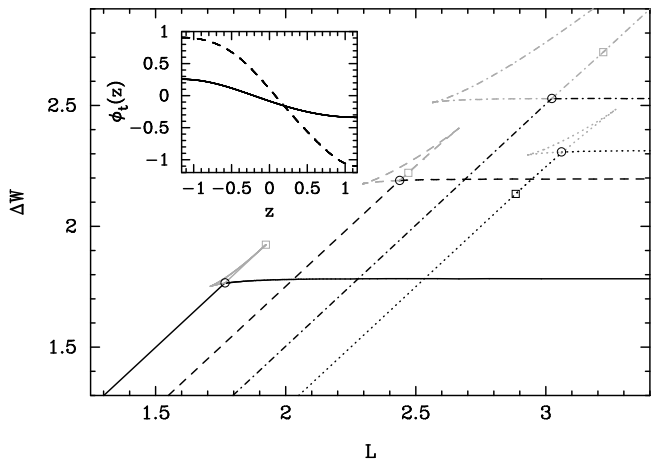} % {firstorder-activation.eps}
  \caption[]{Activation barrier $\Delta W$ (black lines) for various
    potentials exhibiting first-order transitions, marked by open
    circles.  %The line styles correspond to those of Fig.\ \ref{fig:firstorder}.  
    The line styles correspond to the following potentials: 
    (solid) $V(\phi)=1-4\phi^2/3-\phi^4+4\phi^5/3$, 
    (dashed) $V(\phi)=1-\phi^2-\phi^3+\phi^5$, 
    (dot-dashed) $V(\phi)=1-2\phi^2/3-3\phi^3/2+7\phi^5/6$, 
    and (dotted) $V(\phi)=1-13\phi^2/12-\phi^3+\phi^4/4+5\phi^5/6$, 
    with the curves shifted horizontally to
    improve readability. The (second-order) transition between
    uniform and instanton-type escape is marked by an open square for
    each curve.  Activation barriers for higher energy states are
    shown with gray lines.  Degenerate instanton-like saddle states
    corresponding to the first-order transition for the dotted line
    are shown in the inset. Both states have the same length and
    energy. Note that the potential described by the rightmost curve
    (dotted line) exhibits a second-order followed by a first-order 
    transition.  }
  \label{fig:firstorderDW} % 
\end{figure}

Because of the symmetry of the potential (which is not necessary for the
transition to occur~\cite{Stein04}), the change in activation behavior
arises from a {\it bifurcation\/} of the transition state.  Below a
critical length $L_c$ the transition state $\phi_t$ is constant, while
above $L_c$ it becomes a pair of degenerate, spatially varying instanton
configurations~\cite{MS01b}:
\begin{equation}\label{eq:instanton}
  \phi_t   = \begin{cases} 0, & L<L_c  \\ %\label{eq:uniform} \\
	         \pm\sqrt{\frac{2m}{1+m}}{\rm sn}({\frac{2z}{\sqrt{m+1}}}\mid m), \qquad & L\ge L_c %\label{eq:bounce}
	         \end{cases}
\end{equation}
where ${\rm sn}(\cdot\mid m)$ is the Jacobi elliptic $\rm sn$ function with
parameter~$0\le m\le 1$.  Its quarter-period is given by~${\bf K}(m)$, the
complete elliptic integral of the first kind~\cite{Abramowitz65}, which is
a monotonically increasing function of~$m$.  As~$m\to 0^+$, ${\bf K}(m)$
decreases to~$\pi/2$, and ${\rm sn}(\cdot\mid m)\to\sin(\cdot)$.  In this
limit the saddle state smoothly degenerates to the $\phi_t=0$
configuration.  As $m\to 1^-$, the quarter-period increases to infinity
(with a logarithmic divergence), and ${\rm sn}(\cdot\mid
m)\to\tanh(\cdot)$, the (nonperiodic) single-kink sigmoidal function.  The
Langer-Callan-Coleman~\cite{Langer67,Langer69,Coleman77,CC77} `bounce'
solution is thereby recovered as $L\to\infty$.

The value of~$m$ in~(\ref{eq:instanton}) is determined by the interval
length $L$ and the Neumann boundary conditions, which require that
\begin{equation}\label{eq:bc} % boundary conditions
  L = \sqrt{m+1}\,{\bf K}(m)\, 
\end{equation}
The critical length is determined by~(\ref{eq:bc}) when $m=0$; that is,
$L_c=\pi/2$.  
As previously noted, $m\to 1$ corresponds to $L\to\infty$, and
the activation energy smoothly approaches the asymptotic value of
$\Delta W_\infty=4\sqrt2/3$.  
The transition state for an intermediate value of $m$,
corresponding to $L=5$, is shown in the inset of Fig.~\ref{fig:acten}.

% \begin{figure}
%   \includegraphics[width=0.95\columnwidth]{neumannsol.eps}
%   \caption[The activation energy.]{
%     \label{fig:neumannsol} % Solution with Neumann BC
%     The transition  state $\phi_t(z)$ for $L=5$ (corresponding to $m=0.986$) 
%     described by  Eq.~(\ref{eq:bounce}). 
%     Only one of the  symmetric pair is shown.}
% \end{figure}

The activation energy $\Delta W$ can be computed in closed form for all $L>L_c$ 
(below $L_c$, it is simply $\Delta W=L$):
\begin{equation}\label{eq:acten}  % Activation energy
  %\Delta W={1\over3(1+m)^{3/2}}\Bigl[4(1+m){\bf E}(m)-{1\over
  %2}(1-m)(3m+5){\bf K}(m)\Bigr]\, ,
  \Delta W = \frac{8(1+m){\bf E}(m) - (1-m)(3m+5){\bf K}(m)}{3(1+m)^{3/2}}, 
\end{equation}
with ${\bf E}(m)$ the complete elliptic integral of the second
kind~\cite{Abramowitz65}.  
The activation energy as a function of $L$ is shown in Fig.~\ref{fig:acten}(a).  
Note that the curve of $\Delta W$ vs.~$L$
and its first derivative are both continuous at $L_c$; the second
derivative, however, is discontinuous, as might be expected of a
second-order-like phase transition.

A more profound manifestation of critical behavior at $L_c$ is exhibited by
the rate prefactor $\Gamma_0$, which (in the asymptotic limit $T\to 0$)
diverges at $L_c$, as shown in Fig.~\ref{fig:acten}(b).
This is striking, but requires interpretation.  
Because it is not relevant to the present discussion, we refer the interested 
reader to~\cite{Stein05}.

This behavior is generic for a whole class of potentials, as described
below, but first-order transitions have also been observed
\cite{BSS05}, leading to a continuous but non-differentiable
activation barrier $\Delta W$ at the transition points. This is illustrated
for various potentials in Fig.\ \ref{fig:firstorderDW}.  The second-order
transition is still present, but is usually not physically observable, as
it happens for higher energy transition states (note, however, the
exception in Fig.~2, dotted line).

\section{Order of the transition}
\label{sec:order} % Order of the transition

\begin{figure}[b] %%%  Figure 3  %%%%%%%%%%%%%%%%%%%%%%%%%%%%%%%%
  \includegraphics[width=0.99\columnwidth]{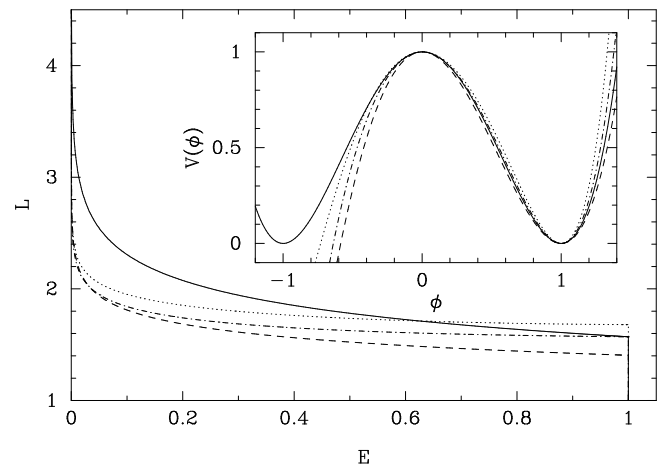} % {secondorder.eps}
  \caption[]{The energy of the classical trajectory determining activation
    behavior for (solid line) the symmetric quartic potential of Eq.~(\ref{eq:quartic}),
    (dashed line) $V(\phi)=1-5\phi^2/2+\phi^3+\phi^4/2$, 
    (dot-dashed line) $V(\phi)=1-2\phi^2+\phi^3/2+\phi^5/2$,
    and (dotted line) $V(\phi)=1-7\phi^2/4+\phi^3/3+5\phi^6/12$.
    The corresponding potentials $V(\phi)$ are plotted in the inset.
    As discussed in the text, the monotonic decrease of energy with interval
    length signifies a second-order phase transition at $L_c$.
    }
  \label{fig:secondorder} % 
\end{figure}

Eqs.~(\ref{eq:variation}) and (\ref{eq:action}) lead to the (typically
nonlinear) differential equation for stationary states
\begin{equation}\label{eq:invpot} % Eq. for stationary state
  \phi''=V'(\phi).
\end{equation}
If we map the field $\phi$ to position and the coordinate $z$ to time, the
solutions to this equation are equivalent to the trajectories of a
classical particle moving in the inverted potential $-V(\phi)$~\cite{Hanggi90}.  
The bounce, or `instanton', transition state (cf.~Eq.\ (\ref{eq:instanton}) 
that determines the activation behavior when $L>L_c$) 
then corresponds to a half-period of such a periodic classical trajectory.

\begin{figure}[t] %%%  Figure 4  %%%%%%%%%%%%%%%%%%%%%%%%%%%%%%%%
  \includegraphics[width=0.99\columnwidth]{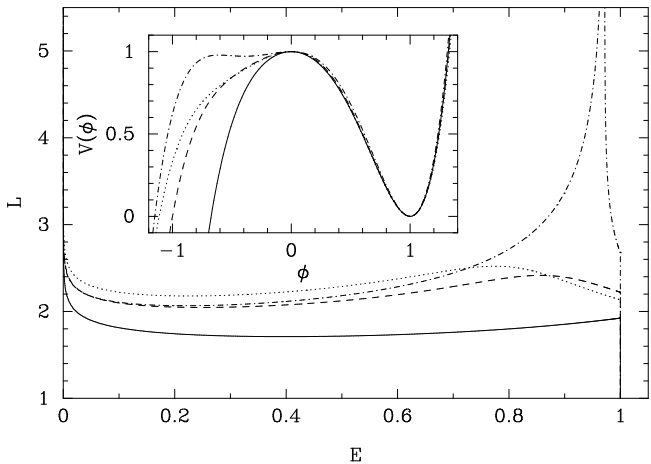} % {firstorder.eps}
  \caption[]{The energy of the classical trajectory determining activation
    behavior for various potentials exhibiting first-order transitions,
    as evidenced by the non-monotonic behavior of $L(E)$.
    The line styles correspond to those of Fig.\ \ref{fig:firstorderDW}, 
%     The curves correspond to the following potentials: 
%     (solid) $V(\phi)=1-4\phi^2/3-\phi^4+4\phi^5/3$, 
%     (dashed) $V(\phi)=1-\phi^2-\phi^3+\phi^5$, 
%     (dot-dashed) $V(\phi)=1-2\phi^2/3-3\phi^3/2+7\phi^5/6$, 
%     and (dotted) $V(\phi)=1-13\phi^2/12-\phi^3+\phi^4/4+5\phi^5/6$.
    with the corresponding potentials $V(\phi)$ plotted in the inset.
    }
  \label{fig:firstorder} % 
\end{figure}

These classical trajectories have a corresponding ``energy'' $\tilde{E}\equiv-E(L)$ given
by (cf.~Eq.~(6) of~\cite{Chudnovsky92})
\begin{equation}\label{eq:energy}
  - E(L) = \frac12(\phi')^2 - V(\phi)\, .
%   \frac12(\phi')^2=V(\phi) - E(L)\, .
\end{equation}
This energy of a classical instanton trajectory should not be confused with
the activation barrier $\Delta W$ given by the action difference between
the transition and metastable states.  $\tilde{E}$ corresponds to the energy of a
classical particle undergoing periodic motion in the inverted potential
$-V(\phi)$.  It is determined either by the temperature in the thermally assisted
tunneling problem, or by the interval length in a stochastic classical
Ginzburg-Landau field theory.

\begin{figure*}[bth] %%%  Figure 5  %%%%%%%%%%%%%%%%%%%%%%%%%%%%%%%%
  \newlength\figfivewidth\setlength\figfivewidth{13.0cm}
    \begin{minipage}[c]{\figfivewidth}\hspace*{-6mm}
       \includegraphics[angle=-90,width=\figfivewidth]{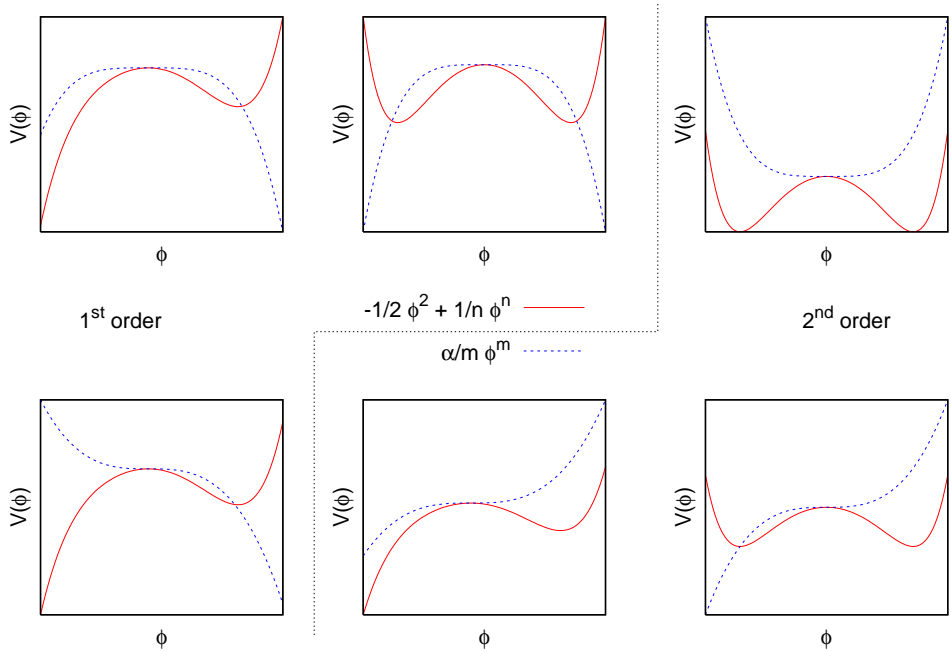} % {PotentialType}
    \end{minipage}\
    \begin{minipage}[c]{4.2cm}
      \caption[]{(Color online) Various types of potentials of the form given by
	Eq.~(\ref{eq:regularpotential}), with the second term $\alpha\phi^m/m$
	plotted as a (blue) dashed line, and the other terms,
	$-\phi^2/2+\phi^n/n$, plotted as a (red) solid line.  Potentials on
	the left-hand-side of the dotted line have first-order transitions,
	while those on the right-hand-side have second-order transitions.
      }\label{fig:TransitionType} % Various types of potentials
    \end{minipage}\vspace*{-3mm}
    
\end{figure*}

We illustrate the behavior of the energy for the simple case of the
symmetric quartic potential given by Eq.~(\ref{eq:quartic}).  For $L<L_c$,
the transition state $\phi_t=0$ [cf.~Eq.\ (\ref{eq:instanton})], and the
energy $E=-\tilde{E}=1$, independent of length. For $L>L_c$, the energy $E$
monotonically decreases to zero as interval length increases.  As a
function of $m$ (related to $L$ through Eq.~(\ref{eq:bc})), it is given by
\begin{equation}\label{eq:secondorderenergy}
  E(m) = 1-\frac{4m}{(1+m)^2}.
\end{equation}

The length as a function of energy is shown in Fig.~\ref{fig:secondorder}
for various potentials, plotted in the figure inset, with the solid line 
corresponding the Eq.\ (\ref{eq:quartic}).  It is instructive to
compare this to the behavior of the activation energy $\Delta W$ as a
function of length (Fig.~\ref{fig:acten}). % and~\ref{fig:firstorderDW}).

As first noted by Chudnovsky~\cite{Chudnovsky92}, the order of the
transition is related to behavior of the period $L$ of the instanton
trajectory vs.~its energy $E$: a monotonic decrease, as in
Fig.~\ref{fig:secondorder}, signifies a second-order transition while a
non-monotonic decrease corresponds to a first-order transition, as shown in
Fig.~\ref{fig:firstorder} for a variety of potentials.

Our interest here is in determining the relation between the potential
properties and the order of the transition.  In order to facilitate
comparison between different potential barriers, we rescale $V(\phi)$
to unit barrier height and width, the latter being defined as the
distance between the maximum and minimum of $V$.  Specifically, we
rescale $V(\phi)$ so that $V(0)=1$, $V'(0)=0$, $V''(0)<0$ and
$V(1)=0$, $V'(1)=0$, $V''(1)>0$.  We will generally take the state 
$\phi\equiv 1$ to be metastable; it can decay toward negative values 
of $\phi$ if there is a $\phi_e<0$ such that $V(\phi_e)\leq 0$.
%   The presence of a metastable state
% requires that there be a $\phi_e<0$ such that $V(\phi_e)\leq 0$.

We begin with some general considerations, valid for all potential
barriers.  All $L(E)$ curves in both Figs.\ \ref{fig:secondorder} and
\ref{fig:firstorder} diverge as $E\rightarrow0^+$.  This generic
behavior is easily understood from the classical-particle analogy, as
a particle with a periodic orbit will have an arbitrarily long period
if its energy is close to, but lower than, a local maximum of the
potential $-V(\phi)$.

Similarly, a barrier containing a local, secondary maximum/minimum pair
(see, e.g., the dot-dashed line in the inset of Fig.\ \ref{fig:firstorder})
will have a divergence of $L(E)$ at the energy corresponding to the
secondary minimum.
% Any local minimum of $V(\phi)$ within the barrier similarly leads to a 
% divergence of $L(E)$.
Thus, a potential barrier that includes a local metastable state leads to a
first-order transition of the activation behavior.  That transition, in
this case, is between a two-step escape through the local minimum and a
direct escape, with both paths having instanton-like transition states.
% An example of such a potential is given by the dot-dashed lines in 
% Figs.\ \ref{fig:firstorderDW} and \ref{fig:firstorder}, with the transition 
% states shown in the inset of the former.

As $L(E)$ initially decreases for increasing $E$ when $E$ is small, a
sufficient condition for having a first-order transition is that
$\text{d}L/\text{d}E>0$ for $E\rightarrow1^-$.  The corresponding condition
on the potential $V$ can be derived analytically using perturbation theory
around the barrier maximum.

More general conclusions require a numerical determination of $L(E)$.  To
simplify the analysis, but still keep the conclusions general, we consider
smooth potentials of the form
\begin{equation}\label{eq:regularpotential} % Regular potential
  V(\phi)=1-\alpha_2\phi^2+{\alpha_m}\phi^m+{\alpha_n}\phi^n,
\end{equation}
with $2<m<n$.  Rescaling to unit barrier height and width %, so that the
% potential has a local maximum $V(\phi=0)=1$ and a minimum $V(\phi=1)=0$
leaves a single free parameter $-n/(n-m)<\alpha_m<2/(m-2)$, with
\begin{align}
\alpha_2=\frac{n+(n-m)\alpha_m}{n-2},
\intertext{and}
\alpha_n=\frac{2-(m-2)\alpha_m}{n-2}.
\end{align}

Our approach is to %consider the behavior of the function $L(\phi_0)$,
% where $\phi_0$ is determined as follows: 
solve numerically the nonlinear differential equation corresponding to a
particle in an inverted potential $-V(\phi)$ with initial condition
$\phi(-L/2)=\phi_0$, with $0\leq\phi_0<1$, and $\phi'(-L/2)=0$, and with
$L$ the minimal length satisfying Neumann boundary conditions at $z=L/2$.
$E(\phi_0)$ follows from Eq.\ (\ref{eq:energy}), thus providing the
function $L(E)$ in parametric form.
% Then if the resulting function $L(E)$ is monotonically decreasing, 
% the transition is second-order; otherwise it's first-order.  

% Our findings are that the transition is unique and second-order if either 
% $\alpha_m>0$ for any $m, n$, or if $\alpha_m<0$, $m=2k+1$ and $n=2l$, 
% with $k$ and $l$ each positive integers such that $1<k<l$.  
% On the other hand, if $\alpha_m<0$ and both $m$ and $n$ are odd, the 
% transition is first-order.  
% We now describe these results in more detail.

% \section{Summary of results}

Our findings are summarized graphically in Fig.\ \ref{fig:TransitionType},
where $-\phi^2/2+\phi^n/n$ (solid lines) and $\alpha\phi^m/m$ (dashed
lines) are plotted separately.  Potentials having first-order transitions
are shown on the left-hand side of the dotted line, and those with
second-order transitions are on the right-hand side.  All potentials on the
left-hand-side have a negative $\phi^m$ term, which appears to be a
necessary, although not a sufficient, condition for the existence of a
first-order transition.

More specifically, for a potential of the form~(\ref{eq:regularpotential}),
the transition is unique and {\bf second-order} if either of the following
conditions is fullfilled:
\begin{itemize}
  \item $\alpha_m \geqslant 0$;
  \item $\alpha_m < 0$, with $m = 2k+1$ and $n=2l$, $k$ and $l$ being 
  positive integers such that $1<k<l$.
\end{itemize}

The second condition is actually a subset of the first for the symmetric
potential $V(-\phi)$. This confirms in particular that polynomial potentials 
with at most quartic terms exhibit only second-order transitions \cite{Stein04}.

On the other hand, the transition is {\bf first-order} if $\alpha_m < 0$ and
\begin{itemize}
  \item $m=2k$, with $k>1$,
  \item[or]{}
  \item $m=2k+1$ and $n=2l+1$, with $1\leqslant k<l$.
\end{itemize}

% Figure\ \ref{fig:TransitionType}, where $-\phi^2/2+\phi^n/n$ (solid lines)
% and $\alpha\phi^m/m$ (dashed lines) are plotted separately, summarizes the
% various combinations of those two terms, with potentials having first-order
% transitions on the left-hand-side of the dotted line, and those with
% second-order transitions on the right-hand-side.  
If the third term on the RHS of Eq.\ (\ref{eq:regularpotential}) is an even
power of $\phi$, then the transition is first-order if and only if
$\alpha_m < 0$.  If it is an odd power of $\phi$, the transition is
first-order if the third term is an odd power of $\phi$ as
well, but with an opposite sign, so that there is a competition between the
two terms.  The mechanism leading to a first order transition thus seems to
be different depending on the parity of the third term in $V$, as discussed
below.

\section{Potentials with a second-order transition}

Potentials given by Eq.\ (\ref{eq:regularpotential}) have a second-order transition 
in the following cases:
\begin{itemize}
  \item $m=2k$, for $k>1$, and $\alpha_m\geqslant0$;
  \item $m=2k+1$, for $k\geqslant1$, and at least one of the conditions
    $n=2l$ or $\alpha_m \geqslant0$.
\end{itemize}

These potential barriers all look very similar, and have the same generic behavior, 
shown in the inset of Fig.\ \ref{fig:secondorder} for various combinations of $m$, $n$, 
and $\alpha_m$: 
The function $L(E)$ decreases monotonically for $E>0$, and the 
escape energy has a linear part corresponding to a homogeneous transition 
for $L<L_c$, with a second-order transition to instanton-like escape at 
$L_c=\pi/\sqrt{2\alpha_2}$.

The transition length is set by the $-\phi^2$ term in the potential: a
potential $V(\phi)=1-\alpha_2\phi^2$ gives an equation for the transition
state that is equivalent to a harmonic oscillator in one dimension, and
therefore has a transition state whose length $L=\pi/\sqrt{2\alpha_2}$ and
energy $E_u=1$ are independent of the initial value $\phi_0$.  The term
proportional to $\phi^n$ modifies that behavior by increasing the length of
the transition state for a given initial value $\phi_0$ compared to the
inverted quadratic potential.  In the analogy to a particle in a
one-dimensional potential $-V$, it decreases the slope of the potential for
larger $\phi$, proportional to the force felt by the particle, thus
increasing the period of the orbit, until one approaches the maximum of
$-V$, where the orbital period diverges.

\begin{figure}[tb] %%%  Figure 6  %%%%%%%%%%%%%%%%%%%%%%%%%%%%%%%%
  \includegraphics[width=0.99\columnwidth]{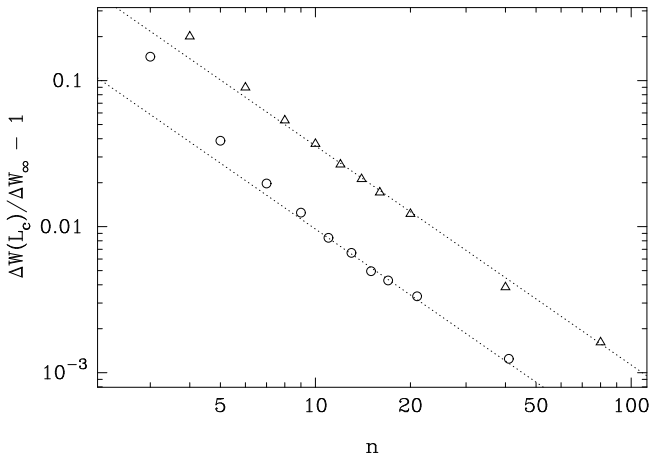} % {EinfovE_Lc}
  \caption{Ratio of the saturation energy barrier $\Delta W_\infty$ to the energy 
    barrier at the transition point $\Delta W(L_c)$ as a function of the power $n$
    of the last term in the potential $V(\phi)=1-\frac{n}{n-2}\phi^2+\frac{2}{n-2}\phi^n$, 
    with circles (triangles) marking odd (even) values of $n$.
    The lines are a guide-to-the-eye showing power-laws with exponenet $3/2$.
    }
  \label{fig:Eratio} % Ratio of saturation and transition point energy barriers
\end{figure}

Considering the case $\alpha_m=0$, one can study numerically the ratio of the 
escape barrier $\Delta W_\infty$ for large interval length $L$ to the energy 
at the transition point $\Delta W(L_c)$ as a function of $n$ 
(see Fig.\ \ref{fig:Eratio}). 
This ratio approaches $1$ as a power law with exponent $3/2$ for $n\gg1$, 
with some even-odd oscillations.
As $n$ becomes large, the transition therefore looks more and more like 
a first-order transition, especially for odd $n$.
Additionally, the curve $L(E)$ remains almost flat for an increasing
range of $E<1$ as $n$ becomes larger.  
This implies that, at the critical length $L_c$, there is a continuum of 
transition states with quasi-degenerate energies available for the escape.

\section{Potential with a first-order transition: Even $m=2k$}

Negative terms of power $m$ larger than 2 in the Taylor expansion of $V(\phi)$ 
around its maximum are responsible for the change of order of the transition, 
but the mechanism creating a first-order transition depends on the parity of $m$.
 
%
% \begin{equation}
%   V(\phi) = 1 -\alpha_2\phi^2 - |\alpha_{k}|\phi^{2k} 
%             + \alpha_n\phi^{n},\qquad \quad 2<2k<n.
% \end{equation}
%

In the case of an even $m=2k>2$, with $\alpha_m=-|\alpha_{2k}|$, 
the function $L(E)$ has a local minimum at some energy $0<E<1$.
% starts decreasing for $\phi_0>0$, reaches a minimum and then increases again. 
The solid line in Fig.\ \ref{fig:firstorder} shows a typical behavior for
that type of potential.  The final increase of $L(E)$ as $E\rightarrow1$,
corresponding to an initial decrease of $L(\phi_0)$ for $\phi_0\ll1$, is
driven by the middle term in $V$, $- |\alpha_{2k}|\phi^{2k}$: on its own,
such a term creates a divergence of the curve $L(E)$ for $E\rightarrow 1$,
or $\phi_0\rightarrow0$, with $L$ thus being an increasing function of $E$.
One can derive an analytic expression for $L(E)$ for a potential
$V(\phi)=1-\alpha\phi^m/m$ by using Eq.\ (\ref{eq:energy}) to express
$\text{d}z$ in terms of $\text{d}\phi$ and $V(\phi)$.  Integrating that
expression yields
\begin{equation}\label{eq:L_E_phim} % L(E) for potential V = 1-phi^m
  L(E)=\frac{\sqrt{2\pi}\,\Gamma(1/m)\alpha^{-1/m}}{m\,\Gamma\big[(m+2)/2m\big]}
    \cdot%\alpha^{-1/m}
      %\left(\frac{m}{\alpha}\right)^{1/m}
      (1-E)^{\frac{2-m}{2m}}.
\end{equation}

% One can show numerically that the `transition state' energy
%
% \begin{equation}
%   E_u(L) = 1-\gamma_m/L^{\beta_m},
% \end{equation}
%
% with $\beta_m=(m+2)/(m-2)$ for $V(\phi) = -\phi^m$.  
The presence of the quadratic term removes the divergence, but the increase
for L(E) as $E\rightarrow1$ remains.  In terms of the classical particle,
the $-\phi^{2k}$ term increases the initial force on the particle compared
to the quadratic case, thus decreasing the oscillation period.  As
$E\rightarrow1^-$, the quadratic term dominates and limits the period to
its value at $L_c$.
% , which still sets the crtical length $L_c$.
% again set by the quadratic term, is the maximum length for the transition 
% state in that potential.
As the energy $E$ decreases the last term in Eq.\ (\ref{eq:regularpotential}) 
inverts the trend and increases the period again, thus creating a minimum in $L(E)$.

This result can be generalized to the potential
\begin{equation}\label{eq:GenPot}
  V(\phi) = 1-\omega^2\phi^2+\phi^{m+2}\sum_{k\geq0}\alpha_k\phi^k,
\end{equation}
with $m>0$ and $\alpha_0\neq0$, for energies $E(\epsilon)=1-\omega^2\epsilon^2$, $\epsilon\ll1$, 
using perturbation theory.  
Using Eq.\ (\ref{eq:energy}), one can derive an expression for the half-period
\begin{equation}\label{eq:LofE}   % Expression for L(E)
  L(E)=\int_{\phi(-\epsilon)}^{\phi(\epsilon)}\frac{\text{d}\phi}{\sqrt{2[V(\phi)-E]}},
\end{equation}
where $\phi(\pm\epsilon)$ is the smallest $|\phi|$ on the right/left-hand-side of the maximum 
such that $V(\phi(\pm\epsilon))\equiv E(\epsilon)$.  
Using a series expansion in $\epsilon$ for $\phi(\pm\epsilon)$, and expanding Eq.\ (\ref{eq:LofE})
to order $\epsilon^{2m}$, one obtains
\begin{multline}\label{eq:LofEseries}  % Series expansion of L(E)
  L(E=1-\omega^2\epsilon^2) = \\
    \frac{\pi}{|\omega|}\left(
    1+\sum_{k=0}^m\frac{1+(-1)^{m+k}}{2}\cdot\frac{\alpha_k}{\omega^2}{\cal A}_k\epsilon^{m+k}\right. \\
     \left.+\frac{\alpha_0^2}{\omega^4}{\cal B}_{2m}\epsilon^{2m} + {\cal O}(\epsilon)^{2m+1}\right),
\end{multline}
where the numbers ${\cal A}_k>0$, $k=1,\dots m$ and ${\cal B}_{2m} >0$.
From Eq.\ (\ref{eq:LofEseries}) it is clear that 
$\text{d}L/\text{d}E=-1/(2\omega^2\epsilon)\cdot\text{d}L/\text{d}\epsilon > 0$ for $E\rightarrow 1$ 
if the lowest even-power $2n>2$ term in the potential $V(\phi)$ has a negative coefficient,
regardless of the presence of odd-power terms, which do not contribute to the slope of $L(E)$ 
to that order, provided $n \leq m$.

This provides a sufficient, though not necessary, condition for the presence of a first-order 
transition: the transition is first-order if the first even power of $\phi$ 
(excluding the quadratic term) in the Taylor expansion of the potential around its maximum 
has a negative coefficient, provided its exponent is at most $2m$, with $m$ defined in 
Eq.\ (\ref{eq:GenPot}).

% Although the divergence is not present for the full potential, the initial
% decrease of $L(\phi_0)$ can be traced back to it.  
% At some point, the last term in $V$, $+ \phi^n$, becomes large enough to 
% invert this trend and make $L$ increase again with increasing $\phi_0$. 
% This point seems to be close to the inflexion point of the potential, 
% although it is not exactly at it.

\section{Potential with a first-order transition: Odd $m=2k+1$}

Perturbation theory yields no information about the influence of odd-power
terms in the potential as they do not contribute (up to order ${\cal
O}(\epsilon)^{2m}$) to the slope of $L(E)$ for $E\rightarrow1^-$.  However
even a negative slope does not imply a second-order transition, as
illustrated in Fig.\ \ref{fig:firstorder} (dotted and dashed lines).  In
order to study the effect of odd-power terms, we return to numerics, with a
potential
\begin{equation}
  V(\phi) = 1 -\alpha_2\phi^2 - |\alpha_k|\phi^{2k+1} 
        + \alpha_l\phi^{2l+1}, %\qquad 1\leq k<l.
\end{equation}
with $1\leq k<l$.

In this case, the function $L(E)$ decreases for $E\rightarrow1^-$, but has
both a maximum and a minimum for lower energies.
% $V$, corresponding to, as $E$ decreases from 1 to 0, respectively a maximum and 
% a minimum of $L(E)$ (see dotted and dashed lines in Fig.\ \ref{fig:firstorder}).  
For $|\alpha_k|>\alpha_{crit}$, the maximum of $L(\phi_0)$ turns into a divergence.  
This is related to the formation of a secondary maximum in the potential $V$, 
as is the case for the dot-dashed line in Fig.\ \ref{fig:firstorder}, 
where $\alpha_k$ is slightly larger than the critical value for that particular potential.  
The maximum of $L(E)$ for $\alpha_k<\alpha_{crit}$ comes from the 
`formation' of that secondary maximum with increasing $\alpha_k$.

Note that the transition from homogeneous to instanton-like escape (marked
by squares in Fig.\ \ref{fig:firstorderDW}, happens above the lowest
activation energy curve (black line) for potentials given by Eq.\
(\ref{eq:regularpotential}), and is therefore not observable.  However the
addition of a positive quartic term to the potential, as is the case for
the dot-dashed line of Figs.\ \ref{fig:firstorderDW} and
\ref{fig:firstorder}, can move that second-order transition %so that it is
`below' the first-order one.  In that case there would be two transitions:
a second-order transition from a homogeneous escape to escape through an
instanton, followed by a first-order transition between two different
escape routes with instanton transition states (shown in the inset of Fig.\
\ref{fig:firstorderDW}).

\section{Discussion}
\label{sec:discussion}

\begin{table}[b]
  \caption{Summary of results: The first column lists the transition order (I or II) 
    for the potential $V(\phi)=-\phi^2+F(\phi)$, with $F(\phi)$ given is the second column.
    The third column lists conditions that $F(\phi)$ needs to satisfy.}
  \label{tab:summary}     %%% Summary of results
  \begin{center}
    \begin{tabular}{lcl}
%       Potential $V(\phi)$ & Conditions \\ %& Transition order \\%[1ex]
%     \hline
%     \multicolumn{3}{l}{Second-order transition for $V(\phi)=-\phi^2 +F(\phi)$ for} \\
      & $F(\phi)$ & Conditions \\
     \hline
%      \multicolumn{2}{l}{I) Second-order} \\
      \rule[2.5ex]{0pt}{0pt}II & $\displaystyle\alpha_m\phi^m+\phi^n$  & $\alpha_m\geq0$, $2<m<n$  \\[1.5ex] %& 2 \\
      II & $\displaystyle\alpha_k\phi^{2k+1}+\phi^{2l}$  & $\alpha_k<0$\\
        & & $2<2k+1<2l$ \\[1ex] %& 2
%     \hline\hline
%     \end{tabular}
    
%     \begin{tabular}{lll}
%     \multicolumn{3}{l}{First-order transition for $V(\phi)=-\phi^2 +F(\phi)$ for} \\
%     B)  & $F(\phi)$ & Conditions \\
%     \multicolumn{2}{l}{II) First-order} \\
    \hline
     \rule[3ex]{0pt}{0pt}I & $\displaystyle\alpha_k\phi^{2k}+\phi^{n}$ & $\alpha_k<0$, $2<2k<n$ \\[1.5ex]
     I & $\displaystyle\alpha_k\phi^{2k+1}+\phi^{2l+1}$ & $\alpha_k<0$, $1<k<l$ \\[1.5ex]
     I & $\displaystyle%\phi^{2m}
          \gamma_n\phi^{2n} +\sum_{k\geq0}\alpha_k\phi^{2(m+k)+1} + \dots$ & $m\geq1$, $n>1$, $n\leq2m$ \\
      & & $\alpha_0\neq0$, $\gamma_n<0$ \\
    \hline
    \end{tabular}
  \end{center}
\end{table}

We have presented a comprehensive study of the dependence of the order of
the barrier crossing transition on potential characteristics, for classical
extended systems subject to weak external spatiotemporal noise.  Using a
combination of analytical and numerical methods, we confirmed an earlier
conjecture of one of the authors~\cite{Stein04} that smooth potentials
whose highest term is quartic have second-order transitions.  We then
considered a wide class of polynomial potentials of arbitrary order, and
determined the potential characteristics that led to either first- or
second-order transitions.  These results are summarized in
Fig.~\ref{fig:TransitionType} and Table~\ref{tab:summary}.  
In particular, we found that the potential
characteristics at the {\it top\/} of the barrier play a central role in
determining the order of the transition.

The order of the transition can play a crucial role in the
understanding of systems near the transition point and in the design
of new experiments (and possibly devices).  For example,
in~\cite{YOT05}, a transition from ohmic to non-ohmic behavior was
observed as the length of the %voltage between 
gold nanowires was changed.  
This was explained~\cite{BSS06} in terms of the transition predicted
in~\cite{BSS05} for monovalent metallic nanowires as wire length
changes.  
% Although the experimental technology is not yet refined
% enough to examine behavior near the transition point, the necessary
% capabilities are likely to arrive within the next few years.
So there already exists experimental evidence for a
transition.  
A more systematic experimental investigation of this ohmic to non-ohmic 
transition is highly desirable in order to examine details of the behavior 
near the transition point.
% to determine the order and nature of the transition in metal nanowires.  
Meanwhile, the analysis given here allows predictions as to which wires 
(characterized by their radius, or more easily measurable, long-wire 
conductance) will undergo one or the other type of transition.
%  a clear prediction can be made: the nature of the transition
% from ohmic to non-ohmic should differ markedly for wires undergoing
% first- vs.~second order transitions.  
These can be measured as a difference
in behavior --- sharp discontinuity vs.~smooth crossover --- of an
effective temperature $T_{eff}(T)$ characterizing the barrier crossing rate
out of a given metastable state~\cite{CG97}.
% In particular, one would expect %a conductivity 
% some linear response functions to have jumps in the former 
% but not the latter (although large fluctuations should be present).  
% The analysis given here allows predictions as to which wires 
% (%as a function of 
% characterized by their radius, or more easily measurable, long-wire 
% conductance) will undergo one or the other type of transition.  
Other applications can be found
in~\cite{Chudnovskybook}.

Although our studies are done in the context of classical transitions
between metastable states, they should be generally applicable to a broad
set of problems, including the classical$\leftrightarrow$quantum crossover
or transitions between regimes of thermally-assisted quantum tunneling,
following the mapping described in~\cite{Stein05}.

\begin{acknowledgments}
This work was supported by NSF Grant Nos.~0312028 (CAS) and PHY-0651077
(DLS).
\end{acknowledgments}

\bibliography{refs}% Produces the bibliography via BibTeX.

\begin{thebibliography}{34}
\bibitem[{not()}]{note}
\bibinfo{note}{The choice of a field dependent on only one spatial dimension is
  primarily for simplification of analysis and discussion. Adding more spatial
  dimensions is unlikely to change the qualitative nature of the transition.
  The essential point is that from the perspective of the phase transition, we
  are already dealing with an infinite-dimensional system, i.e., a field that
  can vary at every spacetime point.}
\end{thebibliography}

\end{document}